\documentclass{ws-procs9x6}
 \def\ksim{\mathrel{\rlap{\lower0.2em\hbox{$\sim$}}\raise0.2em\hbox{$<$}}}
 \def\gsim{\mathrel{\rlap{\lower0.2em\hbox{$\sim$}}\raise0.2em\hbox{$>$}}}
\begin{document}

\title{Quasi-particle model for deconfined matter\footnote{
\uppercase{W}ork supported by \uppercase{BMBF} and 
\uppercase{GSI}.}}

\author{B. K\"ampfer}
\address{Forschungszentrum Rossendorf, PF 510119, 01314 Dresden, Germany}
\author{A. Peshier}
\address{Institut f\"ur Theoretische Physik, Universit\"at Giessen, 
35392 Giessen, Germany}
\author{G. Soff}
\address{Institut f\"ur Theoretische Physik, TU Dresden, 01062 Dresden, Germany}

\maketitle

\abstracts{Our quasi-particle model for deconfined matter near 
$T_c$ is reviewed.
The extrapolation of lattice QCD data to a finite baryo-chemical
potential is discussed.
Determined by the chiral transition temperature $T_c$, the resulting 
equation of state of neutral and $\beta$-stable deconfined matter is 
soft and limits size and mass of pure quark stars.}

\section{Introduction} 

Numerical Monte Carlo evaluations of QCD have succeeded to
deliver the equation of state (EoS) of strongly interacting matter 
(quarks and gluons) with increasing
accuracy. Near and slightly above the chiral transition
temperature (which seems to coincide with the deconfinement temperature)
$T_c$, the EoS shows a non-trivial behavior.
It is, therefore, a challenge to develop models based on QCD,
and catching its relevant degrees of freedom, which describe the EoS
in a quantitative and transparent manner. A big deal of progress has
been made by starting from first principles and arriving, after a chain
of systematic approximations, at models which describe the lattice
QCD data. Despite of several attempts, the range of applicability is
restricted to temperatures $T > 2.5 T_c$.\cite{JPB,Braaten,AP_1} 
Here, we are interested in the range $T_c - 3 T_c$ which covers the
physically important range for heavy-ion physics. We are going to review our
phenomenological quasi-particle model and to describe its use for mapping
lattice QCD data to a finite baryo-chemical potential. In contrast to
the models derived directly from QCD,\cite{JPB,Braaten,AP_1} we use
lattice QCD data to adjust a few parameters. As an application
we discuss implications for cold pure quark stars. 

\section{Quasi-particle model} 

Our model\cite{PLB_94_PRD_96,PRC_00_PRL_01_PRD_02}
is based on the entropy density as combinatoric quantity
measuring the population of states,
$s = \sum_{i = q, g} s_{\rm id}^{(i)} (m_i(T,\mu); T, \mu)$,
which implies for the pressure
$p = \sum_{i = q, g} p_{\rm id}^{(i)} - B(T, \mu)$ via
$s = \partial p / \partial T$. The subscripts ''id'' indicate
the corresponding ideal gas expressions.
All interactions are assumed to be included in the quasi-particle masses 
$m_i^2(T, \mu) \propto \hat \alpha_s$, which
correspond to asymptotic masses of excitations near the light cone
with explicitly given $T$ and $\mu$ dependence from HTL/HDL calculations, 
and the effective coupling strength $\hat \alpha_s (T, \mu)$ is thought 
to include non-perturbative effects beyond that.
Expanding $p(T)$, e.g., for the pure gluon plasma with the lowest order
$m_g^2 = \frac12 g^2 T^2$ one recovers the known perturbative series, 
$p = p_{SB} \left( 1 - \frac{16}{15 \pi^2} g^2
+ O(g^3) + \cdots \right)$.
With the parameterization
$\hat \alpha_s = 12 \pi / (33 - 2 N_f) \ln
\left( \frac{T - T_s}{T_c / \lambda} \right)^2$
we can describe available lattice QCD at $\mu = 0$. Fig.~1 exhibits
two examples.

Due to thermodynamic consistency the trace anomaly is also described correctly.
Opposed to the quasi-particle masses,
the screening mass drops near $T_c$.\cite{PLB_94_PRD_96}
The analysis\cite{temporal} of the temporal quark and gluon propagators 
supports our model.
\begin{figure}[h]
~\\[-.6cm]
\centerline{
\psfig{file=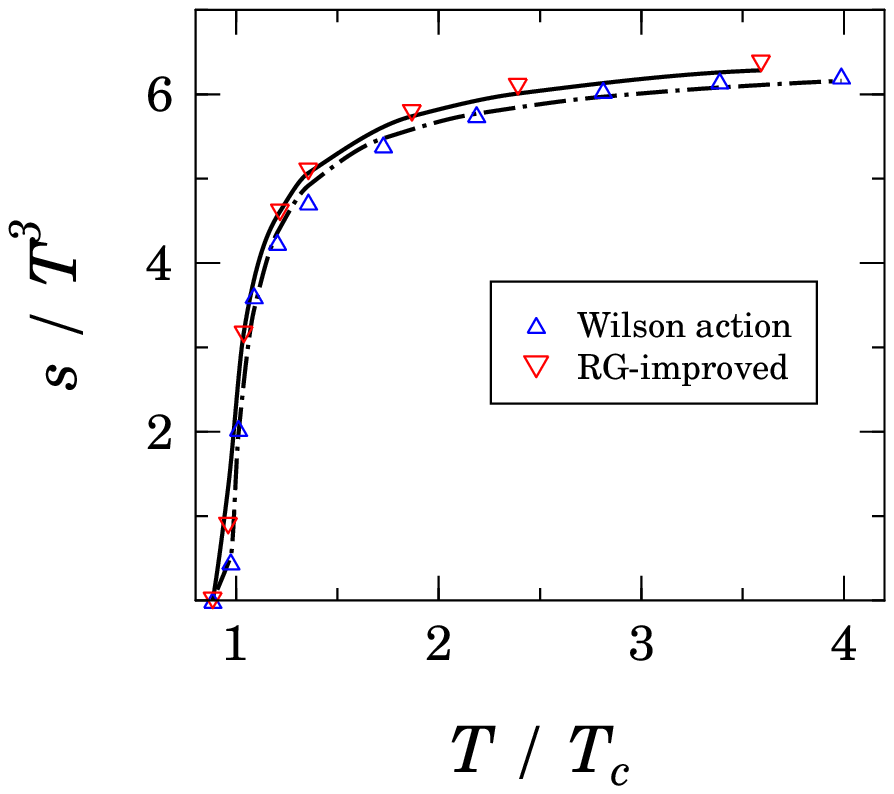,width=5.5cm,angle=-0}
\hfill
\psfig{file=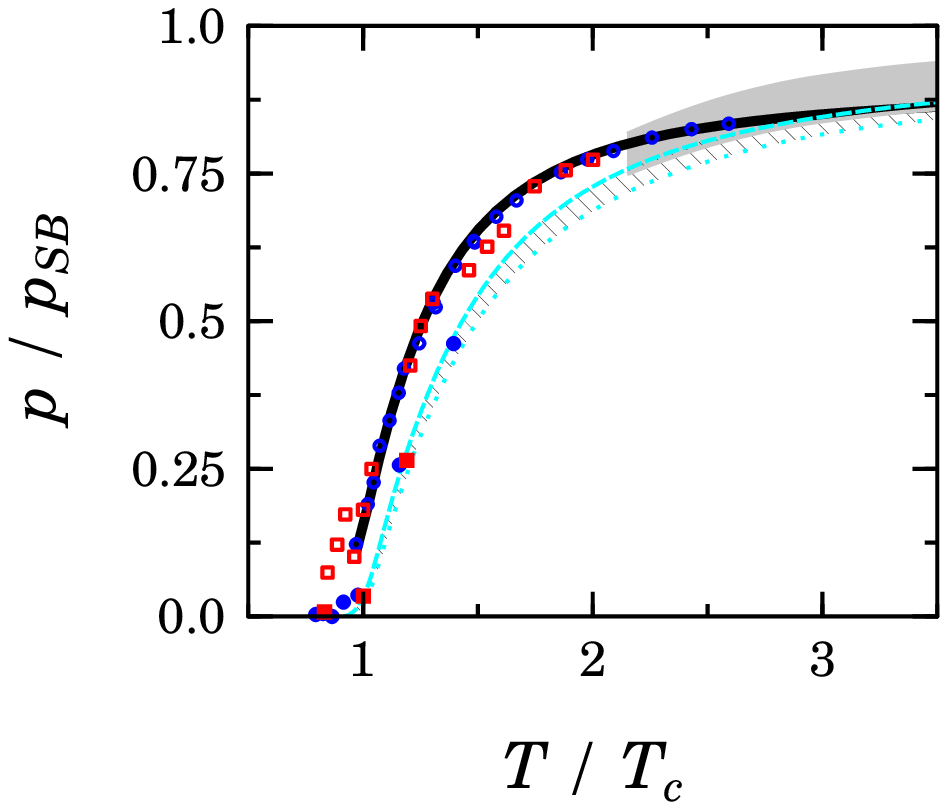,width=5.5cm,angle=-0}} 
\caption{Comparison of lattice QCD results and our quasi-particle model.
Left: pure gluon gas, data from Ref.\protect\cite{Boyed};
Right: $N_f = 2$ quark-gluon plasma, data from Ref.\protect\cite{CP-PACS};
open (closed) symbols are for light (heavier) quarks; the shadowed region
depicts a continuum estimate\protect\cite{Karsch_1}; the hatched area indicates
the pure gluon plasma from the left panel.
The parameters $T_s$, $\lambda$ 
are adjusted to lattice QCD data.
For further details
see Ref.\protect\cite{PRC_00_PRL_01_PRD_02}.  
\label{fig_1}}
\end{figure}

Other quasi-particle formulations are conceivable 
(see Refs.\cite{PLB_94_PRD_96,Heinz_Levai} for quotations). 
For example, in Ref.\cite{Schneider_Weise} a different handling
of the quasi-particle masses has been advocated, where the degeneracies are
taken as temperature dependent and adjustable quantities. 
Our formulation is based on the 
$\phi$-derivable formalism, where the entropy density is
a particularly interesting quantity to be considered.\cite{Eur_Phys_Lett}

Ref.\cite{Moore} makes a strong claim that  
the structure of the above quasi-particle formulation
is not conform to the exactly solvable QCD in
the large $N_f$ limit; however, as argued in Ref.\cite{AP_2} this disclaimer
may not be relevant for the present model which is not aimed at covering
this limit of QCD. 

\section{Extrapolation to finite baryo-chemical potential}

The above discussion is constrained to vanishing baryo-chemical potential.
Due to the stationarity condition of the thermodynamical potential,
$\partial p / \partial m_j = 0$, the entropy and baryon density are given by
the sum of the individual quasi-particle contributions,
$s_i = \partial p_i / \partial T$ and
$n_i = \partial p_i / \partial \mu_i$. 
On the other hand, the pressure $p$ must be a thermodynamic potential so that
$\partial s / \partial \mu = \partial n / \partial T$.
This results in a partial differential equation for $\hat \alpha (T, \mu)$,
which can be solved as well-posed boundary problem with
$\hat \alpha (T, \mu = 0)$ adjusted to lattice QCD data. 
A solution by the method of characteristic curves 
is exhibited in Fig.~2 together with the resulting pressure. 

\begin{figure}[h]
~\\[-.3cm]
\begin{minipage}[t]{11.3cm}
\begin{minipage}[t]{5.5cm}
\psfig{file=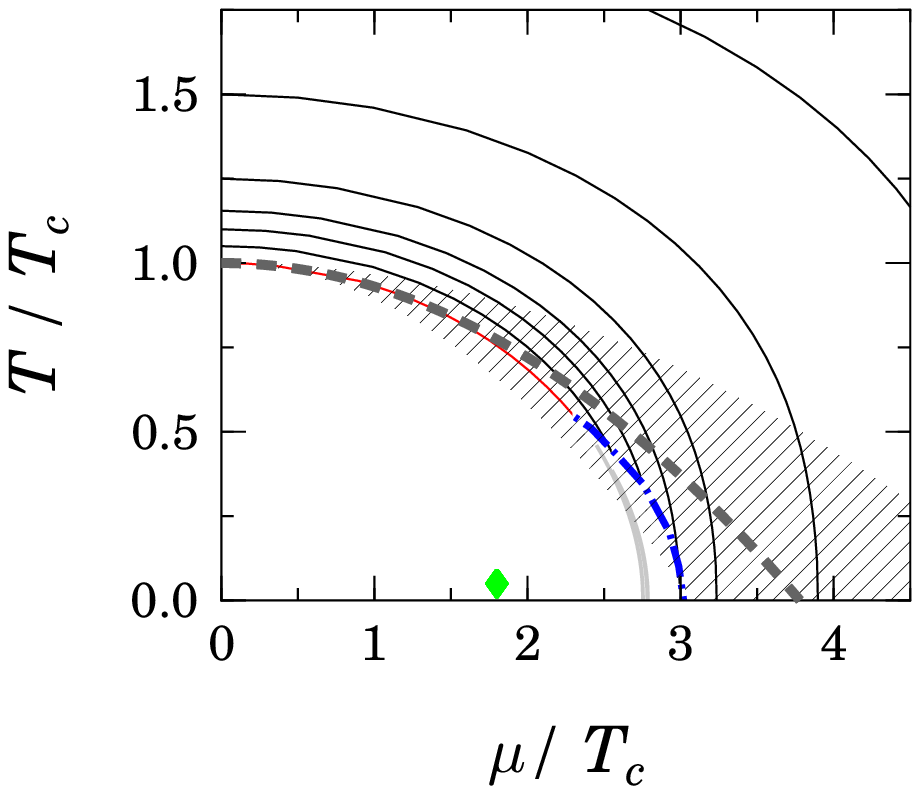,width=5.5cm,angle=-0}
\end{minipage}
\hspace*{9mm}
\begin{minipage}[t]{5.0cm}
\psfig{file=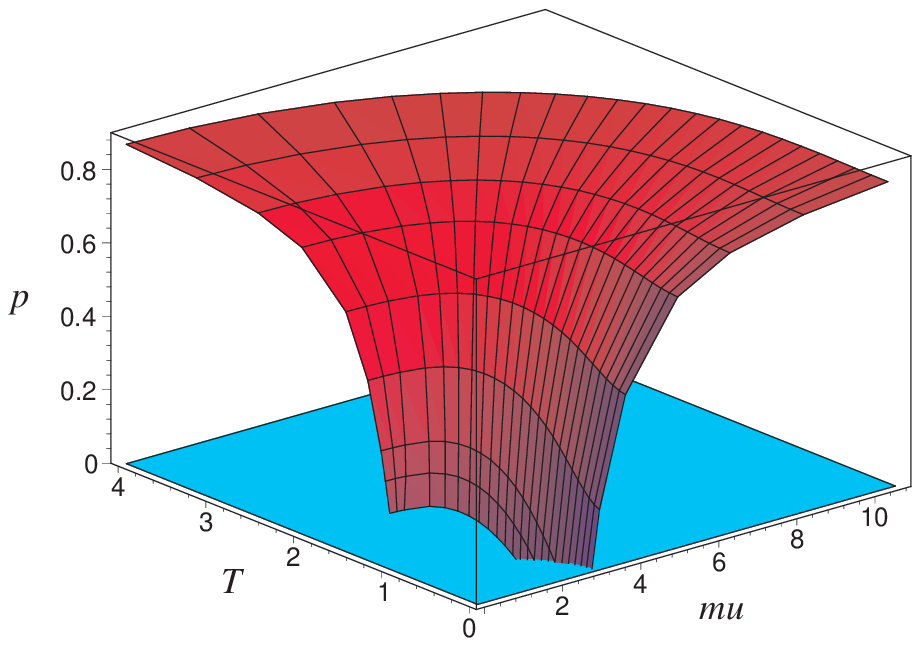,height=4.6cm,angle=-90}
\end{minipage}
\end{minipage}
\caption{Left: The characteristic curves in the $T - \mu$ plane.
The heavy dashed curve is the estimated\protect\cite{swansea} phase border
line (uncertainty indicated by the hatched area). Left to the
dash-dotted curve the pressure becomes negative.
Right: The EoS in our quasi-particle model ($T$ and $\mu$ in units of 
$T_c$, $p$ in units of $p_{SB}$).
Further details in Ref.\protect\cite{PRC_00_PRL_01_PRD_02}.
\label{inter}}
\end{figure}

Very recently, lattice QCD calculations with $\mu > 0$ became 
available.\cite{Fodor} This offers the chance to check the validity of our 
extrapolation of the EoS. A first result is the curvature
of the critical line at $T_c$,\protect\cite{swansea,Fodor}
see Fig.~2. 
Some idea on the robustness of our extrapolation can be gained by
a comparison to the HTL quasi-particle model
applying the same method. The results in Ref.\cite{Romantschek} 
show indeed only a small numerical deviation from our results.  
 
\section{Pure quark stars}   

Soon after the discovery of quarks and gluons as the constituents of
hadrons, the existence of quark cores in neutron stars or pure
quark stars has been conjectured.
Since that time a lot of work has been done to explore various consequences
of this hypothesis.
If the star matter exhibits a first-order phase transition above 
nuclear density, a third family of cold quark stars can appear beyond the 
stable branches of white dwarfs and ordinary neutron stars.\cite{third_island}
The transition, triggered e.g. by accretion, of a neutron star to a
quark star can be accompanied by an ejection of the outermost 
layers.\cite{our_hydro}
This may result in particular supernova events.\cite{SN1987A,Hanauske}
Recently, the possible occurrence of the third family, which would allow 
so-called twin star configurations, has been rediscovered.\cite{Lindblom}
The necessary discontinuity in the EoS may be realized
by the chiral transition, 
but also other phase 
changes of nuclear matter may allow twin stars.\cite{Hanauske}

From the universality of the EoS at $\mu = 0$, the extrapolated
EoS at $\mu > 0$ is rather robust, 
supposed there is no change of the quasi-particle structure.  
All our examined examples deliver approximately
the relation $e = 4 B + \alpha p$ with $\alpha = 3 \cdots 4$ and
$B^{1/4} = \beta T_c$, $\beta = 1.7 \cdots 2.1$
at $T = 0$.\\[-6mm]
\begin{figure}[h]
\begin{minipage}[t]{5.5cm}
~\\[-.7cm]
\psfig{file=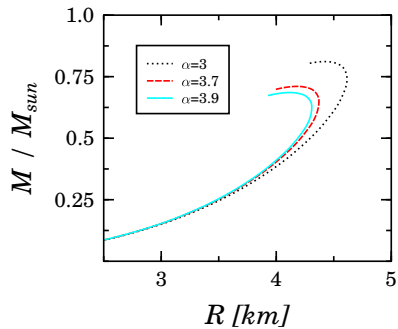,width=5.5cm,angle=-0}
\caption{Mass-radius relation of cold pure quark stars delivered
by our EoS. \label{MR}}
\end{minipage}
\hfill
\begin{minipage}[t]{5.5cm}
Integration of the TOV equations
results in light and small quark stars,\cite{PRC_00_PRL_01_PRD_02}
see Fig.~3. 
Such objects are not (yet?) found.
But one should have in mind, that the outermost layers of our quark star
models are unstable with respect to a phase conversion. Continuing the EoS
in the region where the pressure of our model vanishes by another EoS with
larger pressure can change significantly the mass-radius 
relation.
\end{minipage}
\end{figure}

\section{Summary}

In summary we have reviewed our quasi-particle model of deconfined matter.
To test in more detail the reliability of our model
we still need more information on the EoS
for several flavor numbers with systematic continuum extrapolation and
with realistic quark masses.
Nevertheless, our model allows already at the present stage
an extrapolation of the EoS to cold quark matter. 
The resulting cold pure quark stars
are light and small.
 

\end{document}